\documentclass[10pt,aps,twocolumn,nobibnotes,pra,showpacs]{revtex4-1}
  \usepackage{color}
  \usepackage{newlfont}
  \usepackage{graphicx}
  \usepackage{amssymb}
  \usepackage{amsmath}
  \usepackage{verbatim}
  \usepackage[latin1]{inputenc}
  \usepackage{bbm}
  	
\newcommand{\ket}[1]{\mbox{$ | #1 \rangle $}}
\newcommand{\bra}[1]{\mbox{$ \langle #1 | $}}
\newcommand{\be}{\begin{eqnarray}}
\newcommand{\ee}{\end{eqnarray}}
\begin{document}

\title{Monogamy and backflow of mutual information in non-Markovian thermal baths}
\author{A.~C.~S.~Costa}\email{ana\_sprotte@yahoo.com.br}
\author{R.~M.~Angelo}\email{renato@fisica.ufpr.br}
\author{M.~W.~Beims}\email{mbeims@fisica.ufpr.br}
\affiliation{Department of Physics, Federal University of Paran\'a, P.O.Box 19044, 81531-980, Curitiba, PR, Brazil}

\date{\today}

\begin{abstract}
We investigate the dynamics of information among the parties of tripartite systems. We start by proving two results concerning the monogamy of mutual information. The first one states that mutual information is monogamous for generic tripartite pure states. The second shows that, in general, mutual information is monogamous only if the amount of genuine tripartite correlations is large enough. Then, we analyze the internal dynamics of tripartite systems whose parties do not exchange energy. In particular, we allow for one of the subsystems to play the role of a finite thermal bath. As a result, we find a typical scenario in which local information tends to be converted into delocalized information. Moreover, we show that (i) the information flow is reversible for finite thermal baths at low temperatures, (ii) monogamy of mutual information is respected throughout the dynamics, and (iii) genuine tripartite correlations are typically present. Finally, we analytically calculate a quantity capable of revealing favorable regimes for non-Markovianity in our model.
\end{abstract}


\maketitle

\section{Introduction}


Concepts such as information flow, monogamy, and non-Markovianity have appeared with high frequency in recent literature of quantum information. The reason for that is evident: real-world quantum computers and information processing protocols, as for instance quantum cryptography~\cite{gisin02} and quantum teleportation~\cite{yin12}, invariably depend on how the transfer of information occurs among the numerous constituents of a system and how monogamy and Markovianity constraints this flow.

{Generally speaking, recent efforts have focused on quantifying, characterizing, and controlling information flow in many-body systems. Aiming at understanding the information transfer in condensed-matter systems, Bayat and Bose characterized the ability of different phases of a finite spin chain in transmitting entanglement from an end to another, thus acting as a quantum wire~\cite{bose10}. The dynamics of information in non-Markovian processes has also been investigated in connection with concepts such as quantum Fisher information~\cite{sun10} and geometric phases~\cite{yi10}. Haikka {\em et al} demonstrated how the information flux between an impurity qubit and a Bose-Einstein condensate can be manipulated by engineering the ultracold reservoir with experimentally realistic limits~\cite{maniscalco11}. On the experimental side, an all-optical experiment has been reported which allows one to control the information flow between the system and the environment and to determine the degree of non-Markovianity of the process by measurements on the open system~\cite{piilo11}.} More recently, the flow of quantum correlations in pure states was investigated in three-\cite{walborn14} and multi-partite~\cite{nemes14} systems. By use of the Coffman-Kundu-Wootters (CKW) formula~\cite{wooters00} for the squared concurrence, these works have shown that genuine tripartite entanglement can appear as the coherence initially stored in a given subsystem degrades due to a zero-temperature environment. 

As far as monogamy is concerned, new developments indicate that, while quantum discord {is not generally monogamous~\cite{sen12,fan13}}, the entanglement of formation is as monogamous as concurrence~\cite{fanchini14}. {Interestingly, Fanchini {\em et al} showed, via a monogamic principle, that quantum discord and entanglement of formation define together a conservation law for arbitrary tripartite pure states~\cite{caldeira11}. More recently, Streltsov {\em et al} proved that, in general, only entanglement measures can be strictly monogamous~\cite{brub12}. Nevertheless, Braga {\em et al} were able to derive a monogamy inequality for quantum correlations in a multipartite scenario by showing that the sum of pairwise quantum correlations is upper limited by the global quantum discord~\cite{sarandy12}.} Concerning non-Markovianity, many measures have been proposed~\cite{piilo09,breuer10,plenio10,song12,reza12,paternostro13} and studied for one- and two-qubit systems~\cite{maniscalco13,rossatto13}. For a brief review and a detailed comparison of these measures the reader is referred to Ref.~\cite{maniscalco14}.

In this contribution we link the aforementioned concepts in an approach that extends the above studies to more complex regimes. First, we remove the approximation of zero-temperature reservoirs, which immediately leads us to consider mixed states. Second, we focus our analysis on dephasing dynamics, in which case the information flow is manifestly detached from any energy transfer. Third, we investigate the information dynamics by looking at the total correlations between parties of the system. In particular, we ask under what conditions the mutual information reveals monogamous. Fourth, instead of resorting to the Kraus formalism for arbitrary quantum channels, which is usually employed to model infinite reservoirs, we explicitly consider {\em finite thermal baths}. Such baths may present recurrences regimes which allow us to analyze the backflow of information and non-Markovianity. This study is motivated by the fact that recent technology proved able to access finite environments. As an example we mention the recent observation of single quantum trajectories of a superconducting quantum bit, an achievement that became possible thanks to accurate real-time measurements on the environment~\cite{murch13}. In many problems the thermal bath can be highly structured, containing a finite number of modes which strongly influence back the system dynamics. In fact, the system may be driven towards equilibrium through increasing correlations with the bath \cite{lubkin78,caldeira,gemmer01,zurek03,schlosshauer04,gemmer06,ansgar11,ansgar12}, in contrast with situations in which system and bath remain uncorrelated~\cite{weiss}. Such a complex phenomenon was also observed for the energy transfer between a light-harvesting protein and a reaction center protein \cite{sarovar10,scholes10}. On the theoretical side, decoherence due to finite baths was studied in many works \cite{liu08,paladino02,liang06,wong02,JCP2,frank09,frank10} but remains as a topic of most relevance for chemical physics processes. 

This paper is organized as follows. In Sec.~\ref{info} we make some remarks regarding measures of information. In particular, we show that mutual information is monogamous for mixed states only if genuine tripartite correlations are large enough. In Sec.~\ref{info flow} we study the information flow in dephasing dynamics governed by finite thermal baths. We identify a typical scenario in which local information is converted to monogamous mutual information. A case study is then conducted in Sec.~\ref{case study}, where several results are obtained for the information dynamics. Interestingly, we compute a witness of non-Markovianity and express its behavior as a function of the temperature and the number of modes of the bath. Section~\ref{summary} closes the paper with a summary of our findings.

\section{Preliminary remarks \label{info}}

\subsection{State information and mutual information}
For a generic multipartite system in a state $\rho \in \mathcal{H}$ ($\dim\mathcal{H}=d$) we define the {\em state information} as
\be
\mathcal{I}=\ln d-S,
\label{IS}
\ee
where $S=S(\rho)$ is the von Neumann entropy. Sometimes called {\em negentropy}, $\mathcal{I}$ has been given an operational interpretation in terms of the amount of thermodynamic work that can be extracted from a heat bath when the system state is $\rho$~\cite{barbara05}. Alternatively, it can be viewed as a measure of purity. Consider an arbitrary cut yielding two parties $x$ and $y$ such that $d=d_xd_y$.  It follows from Eq.~\eqref{IS} that
\be
\mathcal{I}=\mathcal{I}_x+\mathcal{I}_y+I_{x:y},
\label{I}
\ee
where $\mathcal{I}_{x}=\mathcal{I}(\rho_{x})$ and $\rho_{x}=\text{Tr}_{y}\rho$. $I_{x:y}=S_x+S_y-S$ is the mutual information of parties $x$ and $y$. From the nonnegativity of the mutual information it follows that $\mathcal{I}\geqslant \mathcal{I}_x+\mathcal{I}_y$, a monogamy relation showing that the total local information is not enough in general to account for the state information; the difference is the mutual information.

In this paper we will focus on tripartite states,  associated with subsystems $A$, $B$, and $C$. Consider that $x=AB$ and $y=C$. According to Eq.~\eqref{I} one can write $\mathcal{I}=\mathcal{I}_{AB}+\mathcal{I}_C+I_{AB:C}$ and $\mathcal{I}_{AB}=\mathcal{I}_A+\mathcal{I}_B+I_{A:B}$, so that 
\be
\mathcal{I}=\mathcal{I}_{\text{\tiny LOC}}+I_{A:B}+I_{AB:C},
\label{I_AB_ABC}
\ee 
where $\mathcal{I}_{\text{\tiny LOC}}\equiv \mathcal{I}_A+\mathcal{I}_B+\mathcal{I}_C$ quantifies the total local information. From Eq.~\eqref{IS} we can also show that $\mathcal{I}=\mathcal{I}_{\text{\tiny LOC}}+I_{T}$, where $I_T\equiv S(\rho\,||\,\rho_A\otimes\rho_B\otimes\rho_C)\geqslant 0$ is the total mutual information~\cite{vedral10} and $S(\rho||\sigma)$ is the relative entropy of $\rho$ and $\sigma$. It is clear that $I_T$ measures the amount of information that is {\em not} stored locally. In fact, it can be written as $I_{T}=\tfrac{1}{3}\big(I_{A:B}+I_{B:C}+I_{A:C}+I_{AB:C}+I_{BC:A}+I_{AC:B}\big)$, the sum of the mutual information of all bipartitions of the system.

\subsection{Genuine tripartite correlations}

Bennett {\em et al}~\cite{horodecki11} define $n$-partite correlations as follows: ``A state of $n$ parties has genuine $n$-partite correlations if it is non-product in every bipartite cut.'' Then they show that $n$-partite correlations accordingly defined satisfy a set of reasonable postulates. As noted by Maziero {\em et al}~\cite{zimmer12}, it follows as a logical implication that
\be
I_3\equiv \min\limits_{\text{\tiny $(A\!,\!B\!,\!C)$}} I_{AB:C}=\min\{I_{AB:C},I_{AC:B},I_{BC:A}\}
\label{I3}
\ee 
turns out to be a measure of genuine tripartite correlations, where the minimization is taken over all permutations of $(A,B,C)$. Throughout this paper we employ this measure to quantify genuine tripartite correlations. Rewriting Eq.~\eqref{I_AB_ABC} as $\mathcal{I}-\mathcal{I}_{\text{\tiny LOC}}-I_{A:B}=I_{AB:C}$ and applying the minimization in both sides, we obtain, by Eq.~\eqref{I3}, that
\be
\mathcal{I}=\mathcal{I}_{\text{\tiny LOC}}+I_3+\max\limits_{\text{\tiny $(A\!,\!B\!,\!C)$}} I_{A:B}.
\ee
Since $\mathcal{I}(t)=\mathcal{I}(0)$ for any closed system, this relation implies a trade off for those measures of information. We use this expression to establish our notion of {\em information flow}. It is clear that whenever the local information changes, the sum of tripartite and bipartite correlations has to change by the same amount.

\subsection{Monogamy of mutual information}

Entanglement is a monogamous correlation~\cite{wooters00,fanchini14,winter04}. This means that it cannot be freely shared by distinct parties. An example of monogamy inequality is the CKW formula for the squared concurrence~\cite{wooters00}, $\mathcal{C}_{(AB)C}^2\geqslant \mathcal{C}_{AC}^2+\mathcal{C}_{BC}^2$, which holds for three-qubit pure states. Since the bipartite entanglement does not generally add up to the total entanglement of the parties $AB$ and $C$, there should be some genuine tripartite entanglement $\tau_{ABC}$ such that $\mathcal{C}_{(AB)C}^2=\mathcal{C}_{AC}^2+\mathcal{C}_{BC}^2+\tau_{ABC}$, with $\tau_{ABC}\geqslant 0$. On the other hand, it is well-known that classical correlations are not monogamous. Here we ask whether there exists some monogamy inequality for the mutual information. To assess this question, we manipulate the definition of mutual information to arrive at
\be
I_{AB:C}&=&\big(I_{A:C}+I_{B:C}\big)+\mathfrak{I},
\label{mon}
\ee
where
\be
\mathfrak{I}=S_{AB}+S_{AC}+S_{BC}-S_A-S_B-S_C-S.
\label{frakI}
\ee 
The classical counterpart of $\mathfrak{I}$---sometimes called {\em interaction information}---appeared long ago in information theory~\cite{McGill,Fano} but, to the best of our knowledge, its interpretation is still debatable~(see \cite{klaus09} and references therein). 

Equation~\eqref{mon} shows that a monogamy inequality will exist if $\mathfrak{I}\geqslant 0$. It has been recently shown by Hayden {\em et al}~\cite{maloney13} that the mutual information is monogamous for quantum field theories with holographic duals. However, it is easy to show, by direct evaluation of $\mathfrak{I}$ for some states, that monogamy is not always satisfied (see Appendix~\ref{appendixA}). In what follows, we identify some situations in which mutual information is assured to be monogamous.

\vskip2mm
\noindent {\bf Result 1.}---{\em Mutual information is monogamous for tripartite pure states.} The proof is given as follows. Since $\rho$ is pure, then $S=0$. The Araki-Lieb inequality implies that $S_{AB}=S_{C}$, $S_{AC}=S_B$, and $S_{BC}=S_A$. This immediately implies that $\mathfrak{I}=0$ and
\be
I_{AB:C}=I_{A:C}+I_{B:C},
\ee
which completes the proof. Within the context of the strong subadditivity (SSA) of the von Neumann entropy, a recent work found out the structure of states that satisfy SSA with equality (see Ref.~\cite{winter04_ssa} and references therein).  From the above calculations, it follows as a simple exercise (see Ref.~\cite{chuang}) that tripartite pure states saturate both forms of the strong subadditivity, i.e., $S_{AC}+S_{BC}=S_A+S_B$ and $S+S_B=S_{AB}+S_{BC}$.

It is clear from Eq.~\eqref{frakI} that $\mathfrak{I}$ is invariant under permutations of the subsystems, this being an expression of its global feature. Then, one may wonder whether this quantity is somehow related to genuine tripartite correlations. To approach this question we minimize Eq.~\eqref{mon} over all permutations of the subsystems, and obtain by Eq.~\eqref{I3} that
\be
I_3=\min_{\text{\tiny $(A\!,\!B\!,\!C)$}}\big(I_{A:C}+I_{B:C}\big)+\mathfrak{I}.
\label{Ig+frakI}
\ee
Besides relating $\mathfrak{I}$ with tripartite correlations, this expression brings us to our second result concerning monogamy.

\vskip2mm
\noindent {\bf Result 2.}---{\em For a generic tripartite state, mutual information is monogamous if and only if the amount of genuine tripartite correlation is large enough, i.e.,}
\be
I_3\geqslant \min_{\text{\tiny $(A\!,\!B\!,\!C)$}}\big(I_{A:C}+I_{B:C}\big).
\label{Ig>min}
\ee
The proof immediately follows from Eqs.~\eqref{mon} and \eqref{Ig+frakI}. In addition, as a corollary of  result 1, it follows that for tripartite pure states the equality holds in \eqref{Ig>min}.

\section{Information flow in dephasing dynamics \label{info flow}}

When two systems interact, they change both energy and coherence. If one of the systems is a reservoir, with ideally infinite degrees of freedom, then two physical processes take place: relaxation and decoherence. While the former is associated with the irreversible loss of energy, the latter refers purely to the loss of purity (dephasing). Typically, decoherence's time is much smaller than relaxation's, which justifies the interest in nondissipative dynamics. 

Here we consider two noninteracting systems, $A$ and $B$, coupled to a common environment $C$ via the Hamiltonian $H=H_A+H_B+H_C+H_{int}$, where
\be
H_{int}=\Big(V_A\otimes \mathbbm{1}_B+\mathbbm{1}_A\otimes V_B\Big)\otimes V_C
\label{Hint}
\ee 
and $V_{X}$ $(X=A,B,C)$ is an operator acting on $\mathcal{H}_{X}$. To focus on nondissipative dynamics, we demand that $[V_X,H_X]=0$, where $H_X$ denotes the free Hamiltonian of the subsystem $X$. Since $H_X$ is local it does not contribute to the dynamics of correlations and can be omitted. We also assume that $A$ and $B$ share an initially correlated state $\rho_{AB}(0)$, whereas $C$ is in the thermal-equilibrium state $\rho_C=e^{-\beta H_C}/Z$, where $Z=\text{Tr}\,e^{-\beta H_C}$, $T=(k_B\beta)^{-1}$ is the equilibrium temperature and $k_B$ is the Boltzmann constant. 

The dynamics of the system is governed by the propagator $U=e^{-iH_{int}t/\hbar}=U_{AC} U_{BC}$, which yields
\be
\rho(t)=U_{AC}\,U_{BC}\,\,\rho_{AB}(0)\,\,U_{BC}^{\dag}\,U_{AC}^{\dag}\,\,\rho_C,
\ee
where $U_{XC}=e^{-iV_X\otimes V_C t/\hbar}$ $(X=A,B)$. Given the form of the interaction and the fact that $[V_C,\rho_C]=0$, we obtain the reduced states
\be
\rho_{XC}(t)&=& U_{XC}\,\,\rho_{X}(0)\otimes\rho_C\,\,U_{XC}^{\dag}
\label{marginals}
\ee
and $\rho_{C}(t)=\rho_C$, where $\rho_{A,B}(0)=\text{Tr}_{B,A}\rho_{A,B}(0)$. It follows that $S_{XC}(t)=S_{XC}(0)=S_X(0)+S_C(0)$ and $S_{C}(t)=S_C(0)$. Therefore,
\be
I_{X:C}(t)+\mathcal{I}_X(t)=\mathcal{I}_X(0) \qquad (X=A,B).
\label{IXC(t)}
\ee
This relation expresses the notion of information flow: any decrease in the information stored locally in $X$ is accompanied with an increase in the mutual information of $X$ and $C$. In other words, the local information decreases because the reservoir $C$, now correlated with $X$, gets to know about this subsystem, so that by measuring $C$ one can get information about $X$. In addition, because $S(t)=S(0)=S_{AB}(0)+S_C(0)$ one has that $I_{AB:C}(t)=\mathcal{I}_{AB}(0)-\mathcal{I}_{AB}(t)$ and
\be
\mathfrak{I}(t)=I_{A:B}(0)-I_{A:B}(t).
\label{frakI(t)}
\ee
Since the system is closed, the state information~\eqref{I_AB_ABC} is constant and can be shown to be $\mathcal{I}(0)=\mathcal{I}_{\text{\tiny LOC}}(0)+I_{A:B}(0)$. Then, we can rewrite Eq.~\eqref{I_AB_ABC} as
\be
I_{X:Y}(t)+I_{XY:Z}(t)=I_{A:B}(0)+\delta \mathcal{I}_{\text{\tiny LOC}}(t),
\label{I(t)}
\ee
where
\be
\delta \mathcal{I}_{\text{\tiny LOC}}(t)\equiv\mathcal{I}_{\text{\tiny LOC}}(0)-\mathcal{I}_{\text{\tiny LOC}}(t)
\label{dI}
\ee
and $(X,Y,Z)$ assume any permutation of $(A,B,C)$. The above results allow us to construct a picture for the information flow in our dephasing model. First, one sees that $I_{A:C}(0)=I_{B:C}(0)=I_{AB:C}(0)=\mathfrak{I}(0)=I_3(0)=0$. In virtue of the nonnegativity of the mutual information, Eq.~\eqref{IXC(t)} implies that $\mathcal{I}_{X}(t)\leqslant \mathcal{I}_{X}(0)$ $(X=A,B)$, whereas $\mathcal{I}_C(t)=\mathcal{I}_C(0)$. {\em These relations show that the information stored locally will generally decrease as the system evolves in time}, that is $\delta \mathcal{I}_{\text{\tiny LOC}}(t)\geqslant 0$. By Eq.~\eqref{I(t)} we see that, as a consequence, the information reappears between distinct parties as mutual information. In fact, Eq.~\eqref{I(t)} implies that not all bipartite information can vanish simultaneously when $I_{A:B}(0)>0$. Moreover, {\em the sum of bipartite correlations has to increase as the local information decreases}. Finally, if $C$ acts as a typical reservoir, then we expect $I_{A:B}(t)$ to decrease with time, so as to produce, according to Eq.~\eqref{frakI(t)}, $\mathfrak{I}(t)>0$. This ensures that throughout the dynamics the mutual information is monogamous. Furthermore, by Eq.~\eqref{Ig+frakI} we see that $I_3(t)>0$ as well. To sum up, our dephasing model reveals a typical scenario in which i) {\em local information transforms to mutual information} and ii) {\em genuine tripartite correlations emerge and ensure monogamy.}

Consider for a while the case in which $\rho_{AB}(0)$ is a state with maximally mixed marginals, i.e., $\rho_{X}(0)=\mathbbm{1}_X/d_X$ $(X=A,B)$. It follows from Eq.~\eqref{marginals} that $\rho_{XC}(t)=\rho_{XC}(0)=\rho_X(0)\otimes\rho_C=\mathbbm{1}_X/d_X\otimes\rho_C$. In this case, there is no dynamics of local information, i.e., $\delta\mathcal{I}_{\text{\tiny LOC}}(t)=0$. Also, it is clear from Eq.~\eqref{IXC(t)} that $I_{X:C}(t)=0$. Hence, by Eqs.~\eqref{Ig+frakI} and \eqref{frakI(t)} we conclude that
\be
I_3(t)=\mathfrak{I}(t)=I_{A:B}(0)-I_{A:B}(t).
\label{I3(t)}
\ee
This result provides a simple way to track the dynamics of genuine tripartite correlations. Moreover, it tells us that $I_3(t)$ will typically increase in ideally infinite thermal baths. On the other hand, as far as finite baths are concerned, the increase in $I_3(t)$ may not be monotonic. In fact, one might suspect that due to eventual information backflow, the tripartite correlations would temporarily decrease. This speculation is assessed in the next section with the aid of a specific model.

\section{Case study: two qubits in a finite bosonic reservoir \label{case study}}

Now we study the information dynamics using a model that allows for an analytical analysis free from usual simplifications, such as the approximations of Born (weak coupling) and Markov (no memory effects). We model the finite bath as a set of $N$ uncoupled harmonic modes, with free Hamiltonian
\be
H_C = \hbar\sum_{k=1}^N \omega_k\hat n_k,
\label{HC}
\ee
where $\omega_k$ is the frequency of the $k$-th mode. The Hamiltonian of the coupling is constructed as in Eq.~\eqref{Hint} with 
\be
V_X = \hbar\,g_X\,\sigma_3^X \qquad \text{and} \qquad V_C=\sum_{k=1}^Ng_k\hat n_k,
\label{VX}
\ee
where $g_X$ is the coupling constant of the qubit $X$ with the bath, $\hat n_k$ is the number operator of the $k$-th mode, and $g_k$ is the coupling constant between the mode $k$ and the qubits. The two qubits are subjected to the same bath, but the strength of the interaction is controlled by $g_X$ $(X=A,B)$. The effectiveness of this bosonic bath in yielding nondissipative decoherence was demonstrated in Ref.~\cite{renato06}.

The two-qubit system is assumed to be initially prepared in a Bell-diagonal state with three real parameters. In the Bloch representation, this state is written as $\rho_{AB}(0)=\rho_{\mathbf{c}}$ where
\be\label{rhox}
\rho_{\mathbf{c}} = \frac{1}{4}\left(\mathbbm{1}_A\otimes\mathbbm{1}_B + \sum_{i=1}^3 c_i \sigma_i^A\otimes\sigma_i^B\right).
\ee
The reduced states are given by $\rho_X=\mathbbm{1}_X/2$. It follows from the analysis carried out in previous section that the generation of genuine tripartite correlations can be quantified via Eq.~\eqref{I3(t)}. Another interesting feature of this model is that the dynamics confines the two-qubit system to a particular subspace. Specifically, the dynamics maps Bell-diagonal states onto Bell-diagonal states. In fact, by computing the time-evolved state $\rho(t)$ and tracing out the reservoir we arrive (using the computational basis) at
\be\label{rhoxt}
\rho_{AB}(t) =
&&\left(
\begin{array}{cccc}
\alpha & 0 & 0 & \delta(t)\\
0 & \beta & \gamma(t) & 0 \\
0 & \gamma^\ast (t) & \beta & 0 \\
\delta^\ast (t) & 0 & 0 & \alpha
\end{array}
\right),
\ee
where  
\be\begin{array}{lll}
\alpha=\tfrac{1+c_3}{4}, &\hskip3mm & \gamma(t)=\tfrac{(c_1+c_2)}{4}\,\theta_-(t), \\ \\
\beta=\tfrac{1-c3}{4}, & &\delta(t)=\tfrac{(c_1-c_2)}{4}\,\theta_+(t),
\end{array}\ee
and
\be
\theta_{\pm}(t) = \prod_{k=1}^N \left(\frac{1-e^{-\beta\,\hbar\omega_k}}{1 - e^{-[\beta\,\hbar\omega_k
 + 2i(g_A\pm g_B)\,g_kt]}}\right).
\ee
Interestingly, all the influence of thermal bath on the two-qubit system is encoded in the functions $\theta_{\pm}(t)$, which depend on the number $N$ of modes, their frequencies and the coupling parameters ($g_k,\,g_A$, and $g_B$). Although the antidiagonal elements are complex, they can be made real by the local unitary transformation $e^{-i\phi_A\sigma_3^A}\otimes e^{-i\phi_B\sigma_3^B}$~\cite{huang13}, with $\phi_A=-\tfrac{(\varphi_++\varphi_-)}{4}$ and $\phi_B=-\tfrac{(\varphi_+-\varphi_-)}{4}$, where $\theta_{\pm}=|\theta_{\pm}|e^{i\varphi_{\pm}}$. This procedure brings the state \eqref{rhoxt} back to the class of three-parameter Bell-diagonal states, i.e., $\rho_{AB}(t)=\rho_{\mathbf{c}'}$, where $\mathbf{c}'=(c_1',c_2',c_3')$ and
\be\begin{array}{l}
c_1'=c_1\left(\frac{|\theta_-|+|\theta_+|}{2}\right)+c_2\left(\frac{|\theta_-|-|\theta_+|}{2}\right),\\
c_2'=c_1\left(\frac{|\theta_-|-|\theta_+|}{2}\right)+c_2\left(\frac{|\theta_-|+|\theta_+|}{2}\right),\\
c_3'=c_3.
\end{array}\label{ci'}\ee
As a consequence, one has that $\rho_X(t)=\mathbbm{1}_X/2$, which implies that there will be no dynamics of local information. The calculations show that the locally transformed state is identical to \eqref{rhoxt} provided we replace the functions $\theta_{\pm}(t)$ with their moduli,
\be\label{thetamod}
|\theta_\pm (t)| = \prod_{k=1}^N \left(1 + 
\frac{\sin^2 [(g_A \pm g_B)\,g_k \,t]}{\sinh^2 (\beta\,\hbar\omega_k/2)}\right)^{-1/2}.
\label{|theta|}
\ee
Even though general models of baths assume that the frequencies of its constituents obey some distribution, here we will admit, as a simplifying hypothesis, that $\omega_k=\omega_0$. This assumption is justified by the fact---verified numerically---that a distribution for $g_k$ is more effective in causing decoherence than would be a distribution for $\omega_k$ with $g_k=g_0$. For the present analysis we will take a Gaussian spectral distribution for the coupling parameters,
\be
g_k = g_0 \exp\left(-\frac{k^2}{\delta^2}\right),
\ee
where $\delta$ controls the width of the distribution and $g_0$ gives the strength of the coupling.

Now we obtain the main defining features of our reservoir model. First, we compute the decoherence time. To this end, we consider a short time regime, $(g_A \pm g_B)t g_k \ll 1$, which allows Eq.~\eqref{|theta|} to be approximated by
\be
|\theta_\pm (t)| \cong \exp\left[- \sum_k \frac{(g_A \pm g_B)^2 g_k^2\,t^2}{2 \sinh^2(\beta\hbar\omega_0 /2)}\right].
\ee
This result can be rewritten as $|\theta_\pm (t)| \cong e^{-t^2/t^2_D}$, where $t_D$---the decoherence time---is given by
\be\label{tD}
t_{D} = \frac{\sinh (\beta\hbar\omega_0/2)}{g\, G},
\ee
where $g=\min\{|g_A+g_B|,|g_A-g_B|\}$ and $G^2=\tfrac{1}{2}\sum_k g_k^2$. In the limit $N\to \infty$, one may show that $G^2=\sqrt{\tfrac{\pi}{8}}g_0^2\,\delta\,\xi$, where $\xi=\tfrac{\vartheta_3(0,\exp(-2/\delta^2))-1}{\sqrt{2\pi\delta^2}}\leqslant \tfrac{1}{2}$ and $\vartheta_3$ is a Jacobi theta function. It follows that $G^2\leqslant \sqrt{\tfrac{\pi}{32}}g_0^2\delta$, equality holding strictly for $\delta\to\infty$ and approximately for $\delta \gg 1$. The link with Ohmic environments can be established by taking $G^2=\int_0^{\infty}d\omega J(\omega)$ for a Ohmic-like spectral density $J(\omega)=\eta \tfrac{\omega^s}{\omega_c^{s-1}}e^{-\omega/\omega_c}$, where $\omega_c$ is the cutoff frequency, $\eta$ is a dimensionless coupling constant and $s$ is a parameter that regulates whether the reservoir is sub-Ohmic $(s<1)$, Ohmic $(s=1)$, or super-Ohmic $(s>1)$~\cite{rossatto13}. By performing the integration in $\omega$ we obtain the relation $\sqrt{\tfrac{\pi}{32}}g_0^2\delta=\omega_c^2\eta\Gamma(1+s)$, where $\Gamma$ is the gamma function. Then, for $s>-1$ the identifications $\omega_c^2\propto \delta$ and $\eta \propto g_0^2$ allow us to simulate an Ohmic bath. As far as we consider $N$ finite, however, we can check that 
\be
G\cong\tfrac{g_0}{2}\left(\sqrt{\tfrac{\pi}{2}}\,\delta-1 \right)^{\!1/2}
\ee 
is a rather good approximation for $0\leqslant \delta \leqslant N$. Clearly, decoherence is more destructive when $\delta$ is large. In particular, we can use $\delta\sim N$ (for $N\geqslant 2$), which accentuates the variation in the spectral profile. In this case, since $G\propto N^{1/2}$, the decoherence time scales as $N^{-1/2}$, thus decreasing with the number of modes, as expected. Also, $t_D$ decreases with the temperature. For low temperatures, however, decoherence can still occur, provided that $N$ is large enough. We can qualify the competition between $N$ and $T$ as follows. First we write $|\theta_\pm (t)| = [\prod_k (1 + \frac{x_k^2}{y_0^2})]^{-1/2}$, with
$x_k = \sin [(g_A \pm g_B)t g_k]$ and $y_0 = \sinh (\beta\hbar\omega_0/2)$. Given that $|x_k|\leqslant 1$, one can show that $|\theta_{\pm}(t)|\geqslant |\theta|_{\text{min}}$, where
\be\label{theta_min}
|\theta|_{\text{min}}=\tanh^N(\beta\,\hbar\omega_0/2)\leqslant 1.
\ee
This is the value reached by $|\theta_{\pm}(t)|$ for $t>t_D$. Therefore, strictly speaking, decoherence is not complete for finite $N$. However, it is always possible to make the minimum arbitrarily small by increasing the temperature. 

\begin{figure}[htb]
\includegraphics[width=\columnwidth]{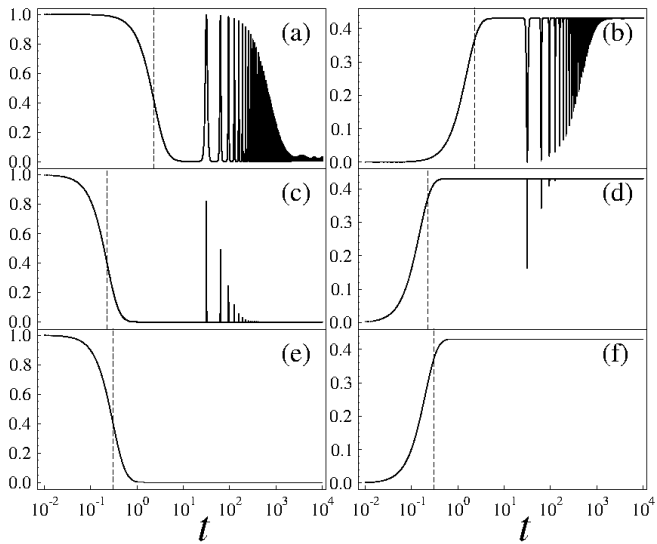}
\caption{$|\theta_-(t)|$ (first column) and $I_3(t)$ (second column) as a function of time (in log scale) for $N=10$, $c_{1,2,3}=-0.8$, $\hbar=\omega_0=1$, $g_A=1$, $g_B=2$, and $g_0=0.1$. The varying parameters are the temperature and the width of the spectral distribution: (a,b) $\beta=1$ and $\delta=10 N$; (c,d) $\beta=0.1$ and $\delta=10 N$; (e,f) $\beta=0.1$ and $\delta=N$. All parameters are given in arbitrary units. The vertical line in each panel accounts for the decoherence time \eqref{tD}.}
\label{thetaI3}
\end{figure}

In Fig.~\ref{thetaI3}, $|\theta_-(t)|$ and $I_3(t)$ are shown as a function of time for a very small thermal bath $(N=10)$. Interestingly, {\em recurrences} occur for low temperatures and a smooth spectral density (panels (a-d)). It is obvious by Eq.~\eqref{I3(t)} that tripartite correlations increase at expense of the correlations between the qubits. Although the simulations shown concern the Werner state $c_{1,2,3}=-0.8$, the scenario there illustrated is typical because the decoherence dynamics is mostly governed by $|\theta_{\pm}(t)|$. The dependence on the initial state reflects only in the amplitude of $I_3$.

From $\rho_X(t)=\rho_X(0)=\mathbbm{1}_X/2$ and Eqs.~\eqref{I(t)} and \eqref{I3(t)} it follows that 
\be
I_{AB:C}(t)=I_3(t). 
\label{IAB:CI3}
\ee
This shows that the tripartite correlations emerge from the flow of the information initially stored in $AB$ to $C$. Also, since $I_{X:C}(t)=0$, the reservoir $C$ knows nothing about $A$ and $B$ individually, only about $AB$. 

\subsection{Quantum versus classical correlations}

\begin{figure}[htb]
\includegraphics[width=\columnwidth]{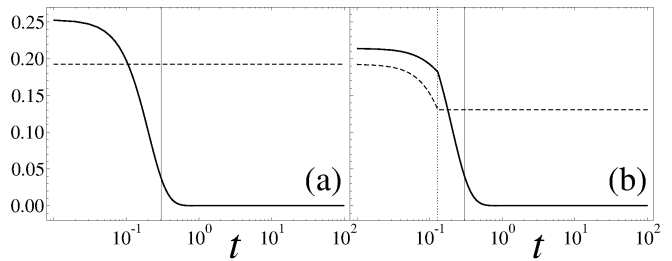}
\caption{Quantum discord $D_{AB}^{\longleftarrow}(t)$ (solid line) and accessible information $J_{AB}^{\longleftarrow}(t)$ (dashed line) as a function of time (in log scale) for the same parameters of Fig.~\ref{thetaI3}-(e,f). In (a) $c_{1,2,3}=-0.6$ and in (b) $c_{1,2}=-0.6$ and $c_3=-0.5$. The {solid} vertical lines account for the decoherence time \eqref{tD}, {whereas the dashed vertical line in (b) accounts for the sudden-change time~\eqref{tPB}}. The constant value of the accessible information is given by $J_{AB}^{\longleftarrow}(|c_3|)$.}
\label{DJ}
\end{figure}

Given that $I_{A:B}(t)=I_{A:B}(0)-I_3(t)$, it is clear that the total correlations between $A$ and $B$ decreases with time. To see what happens with the flow of quantum and classical correlations separately, we compute the quantum discord $D_{AB}^{\longleftarrow}$, whose formula is well known for Bell-diagonal states~\cite{luo08,costa13}), and the accessible information $J_{AB}^{\longleftarrow}=I_{A:B}-D_{AB}^{\longleftarrow}$ (a measure of classical correlations~\cite{vedral01}), which can be show to be~\cite{luo08}
\be
J_{AB}^{\longleftarrow}(c_t)=\frac{(1+c_t)}{2}\ln\left(1+c_t\right)+\frac{(1-c_t)}{2}\ln\left(1-c_t\right),\nonumber
\ee
where $c_t=\max\{|c_1'|,|c_2'|,|c_3'| \}$. From Eq.~\eqref{ci'} we see that $|c_{1,2}'|\leqslant|c_{1,2}|$. Hence, for any initial state such that $|c_3|\geqslant|c_{1,2}|$, we have that $c_t=|c_3|$ and, therefore, $J_{AB}^{\longleftarrow}(t)=J_{AB}^{\longleftarrow}(0)$. In this case, $I_{A:B}(t)=D_{AB}^{\longleftarrow}(t)+J_{AB}^{\longleftarrow}(0)$, which implies, by Eq.~\eqref{I3(t)}, that 
\be
I_3(t) + D_{AB}^{\longleftarrow}(t)=D_{AB}^{\longleftarrow}(0).
\label{I3D}
\ee
This result identifies a class of states for which any increase in the genuine tripartite correlations occurs at expense of the quantum correlations between $A$ and $B$. For states such that $|c_3|<|c_{1,2}|$, the accessible information decreases with time until $|c_3|>|c_{1,2}'|$, a condition that is invariably reached as decoherence takes place. At this stage, the accessible information assumes the constant value $J_{AB}^{\longleftarrow}(|c_3|)$ and the ``conservation relation'' \eqref{I3D} starts to hold (see Fig.~\ref{DJ}). {It was recently shown that the instant $t_{PB}$, at which the accessible information suddenly reaches a constant value, signals the emergence of the {\em pointer basis}, a crucial element in approaches to the measurement problem~\cite{ribeiro12}. In our model, $t_{PB}$ can be analytically computed when $c_2=-\epsilon c_1$ with $\epsilon=\pm 1$, for in this case we have that $|c_{1,2}'|=|c_1||\theta_{\epsilon}|$. Imposing that $|c_{1,2}'|=|c_3|$ and using the short-time approximation for $|\theta_{\epsilon} (t_{PB})|$ we obtain that
\be\label{tPB}
t_{PB}^2=t_D^2\ln\frac{|c_1|}{|c_3|},
\ee
where $t_D$ is to be calculated via Eq.~\eqref{tD} with $g=|g_A+\epsilon g_B|$. Clearly, a sudden transition will occur only if $|c_1|>|c_3|$.
}

We also evaluated the entanglement of formation $E_{AB}$~\cite{eberly07}. The results showed that the entanglement typically undergoes sudden death, which happens soon after the decoherence time, but it can eventually reappear for regimes such as those of Fig.~\ref{thetaI3}--(a-d).

\subsection{A measure of non-Markovianity}

\begin{figure}[htb]
\centerline{\includegraphics[scale=0.22]{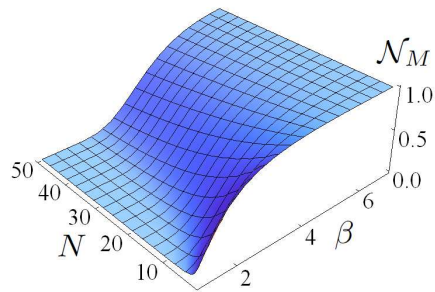}}
\caption{(Color online) Non-Markovianity $\mathcal{N}_M$ as a function of the inverse temperature $\beta$ and the number of modes $N$ for $\hbar=\omega_0=1$, $g_A=1$, $g_B=2$, $g_0=0.1$, and $\delta=10 N$. In this simulation we used $t=10^6$. All parameters are given in arbitrary units. Non-Markovianity is favored by small reservoirs and low temperatures.}
\label{NMfig}
\end{figure}

We now discuss the conditions under which our reservoir model is non-Markovian. To this end, we introduce a witness of non-Markovianity (NM) that is inspired by well-established measures~\cite{piilo09,plenio10,song12} but has the advantage of being easily computable for our model. Here we associate the notion of NM with the capability of the process in allowing for the backflow of correlations from the reservoir to the system. In our model, this is signaled by recurrences in $I_{AB:C}(t)$, which measures how much the reservoir $C$ gets to know about the system $AB$. Then, by Eq.~\eqref{IAB:CI3} and the results shown in Fig.~\ref{thetaI3}, we can conclude that the mechanisms of NM can be investigated directly in $|\theta_{\pm}(t)|$, a state-independent quantity. For simplicity, we replace $|\theta_{\pm}(t)|$ with  $|\theta (t)| = [\prod_k (1 + \frac{x_k^2}{y_0^2})]^{-1/2}$, where $x_k = \sin (g g_k t)$, $y_0 = \sinh (\beta\hbar\omega_0/2)$, and $g=\min\{|g_A+g_B|,|g_A-g_B|\}$. If $y_0\gg x_k$, then $|\theta|\cong e^{-\sum_kx_k^2/2y_0^2}$. This result can be conveniently written as 
\be
|\theta(t)|\cong \exp\left(-\frac{N}{4y_0^2}\big[1-\langle c\rangle(t)\big] \right),
\label{theta_approx}
\ee
where
\be
\langle c\rangle (t) = \frac{1}{N}\sum\limits_{k=1}^N\cos\big(2g g_kt\big).
\ee
Since recurrences in $|\theta(t)|$ are a symptom of NM, we define our measure as
\be
\mathcal{N}_M(t)=\frac{1}{t}\int_{0}^{t}dt' |\theta(t')|.
\ee
This quantity increases with the number of recurrences occurring up to the instant $t$. Using again the limit of low temperatures $(y_0^2\gg N)$ we expand Eq.~\eqref{theta_approx}, perform the time integral analytically, and then turn the result back to the exponential form. The result reads
\be
\mathcal{N}_M(t)=\exp\left\{-\frac{N[\,1-\overline{\langle c\rangle}(t)\,]}{4\sinh^2(\beta\,\hbar\omega_0/2)} \right\},
\label{NM}
\ee
where
\be
\overline{\langle c\rangle}(t)=\frac{1}{N}\sum\limits_{k=1}^N\frac{\sin(2gg_kt)}{2 gg_kt}.
\ee
These expressions hold in the weak-coupling regime. They reveal the conditions for NM to occur. When $g=0$ one has that $\overline{\langle c\rangle}(t)=1$ and the NM of the process is maximum ($\mathcal{N}_M(t)=1$), as expected. This shows that NM is favored by weak coupling. For $g>0$, two time regimes are noticeable. While in the short-time regime NM is influenced by the spectral distribution $g_k$, at the equilibrium this distribution plays no role at all. In fact, it is clear that $\overline{\langle c\rangle}(\infty)=0$ and 
\be
\mathcal{N}_M(\infty)=\exp\left(-\frac{N}{4\sinh^2[\beta\,\hbar\omega_0/2]} \right).
\ee
This formula identifies the physical parameters that crucially influence NM. One sees that NM can be significantly enhanced for small baths ($N$ small) and low temperatures.  This was also observed in classical systems coupled to finite baths \cite{janePRE,janePD09}. For the simulations shown in Fig.~\ref{thetaI3}, our measure results $\ln\mathcal{N}_M=-9.206$ in (a), $\ln\mathcal{N}_M=-999.086$ in (c), and $\ln\mathcal{N}_M=-999.212$ in (e), when computed for $t=10^4$, thus suggesting that the processes in (c) and (e) are strongly Markovian. Although these results agree with the scenario illustrated in the figure, those simulations were obtained in the regime of high temperatures, for which it is not clear whether our measure can give accurate results. In Fig.~\ref{NMfig} we illustrate the behavior of $\mathcal{N}_M$ as a function of the inverse temperature $\beta$ and the number of modes $N$ for a given regime of coupling. 

\section{Summary\label{summary}}
In this paper we conducted a thorough analysis of the information flow in the context of dephasing dynamics induced by finite baths. We started by deriving some results for the main figure of merit in our work, namely, the mutual information. Specifically, we showed that mutual information is monogamous (i) for all tripartite pure states and (ii) for tripartite mixed states for which the amount of genuine tripartite correlations is greater than a certain lower bound. Besides complementing recent studies on the monogamy of correlations~\cite{fan13,maloney13,fanchini14}, our result establishes an interesting link between the mutual information monogamy and tripartite correlations. Concerning dephasing dynamics, we found out a typical scenario in which the information associated to subsystems delocalizes within the system. In addition, our results show that genuine tripartite correlations will generally increase, thus ensuring monogamy for the mutual information. At last, we provided an analytical study for a model of two noninteracting qubits coupled with a nondissipative finite thermal bath. Besides illustrating our predictions for general dephasing models, this case study allowed for the observation of two relevant aspects. First, we verified the existence of a conservation relation involving the amount of tripartite correlations and the amount of quantum correlations in the two-qubit system. Second, we calculated a measure of non-Markovianity which revealed the quantitative dependence of the decoherence process with the number of modes of the thermal bath, the equilibrium temperature, and the spectral distribution.

\section*{Acknowledgments}

This work was supported by CAPES, CNPq, and the National Institute for Science and Technology of Quantum Information (INCT-IQ). The authors thank W. T. Strunz for fruitful discussions.

\appendix
\section{Genuine tripartite correlations and interaction information for three-qubit states\label{appendixA}}

\begin{figure}[htb]
\centering
\includegraphics[scale=0.168]{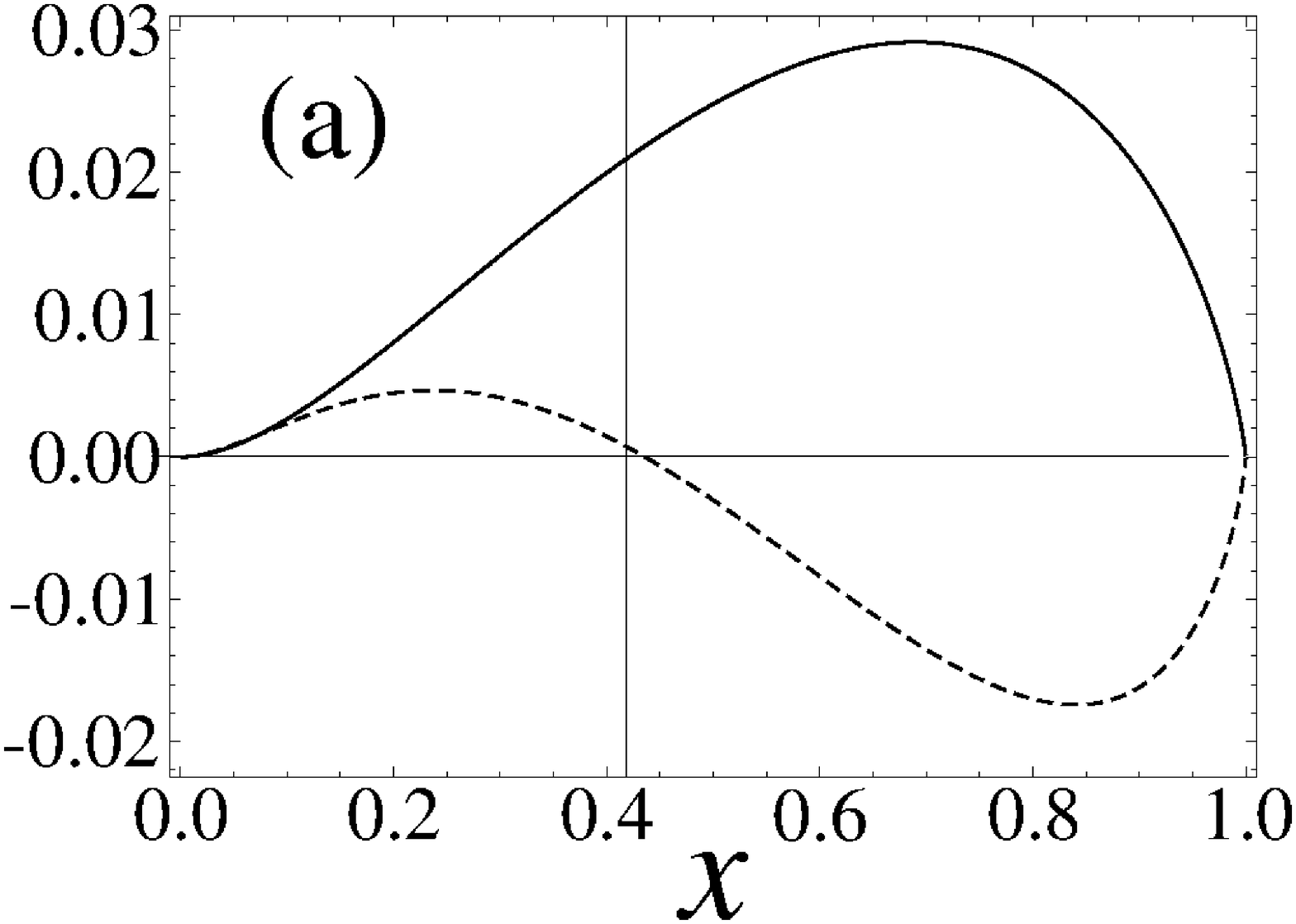}\includegraphics[scale=0.168]{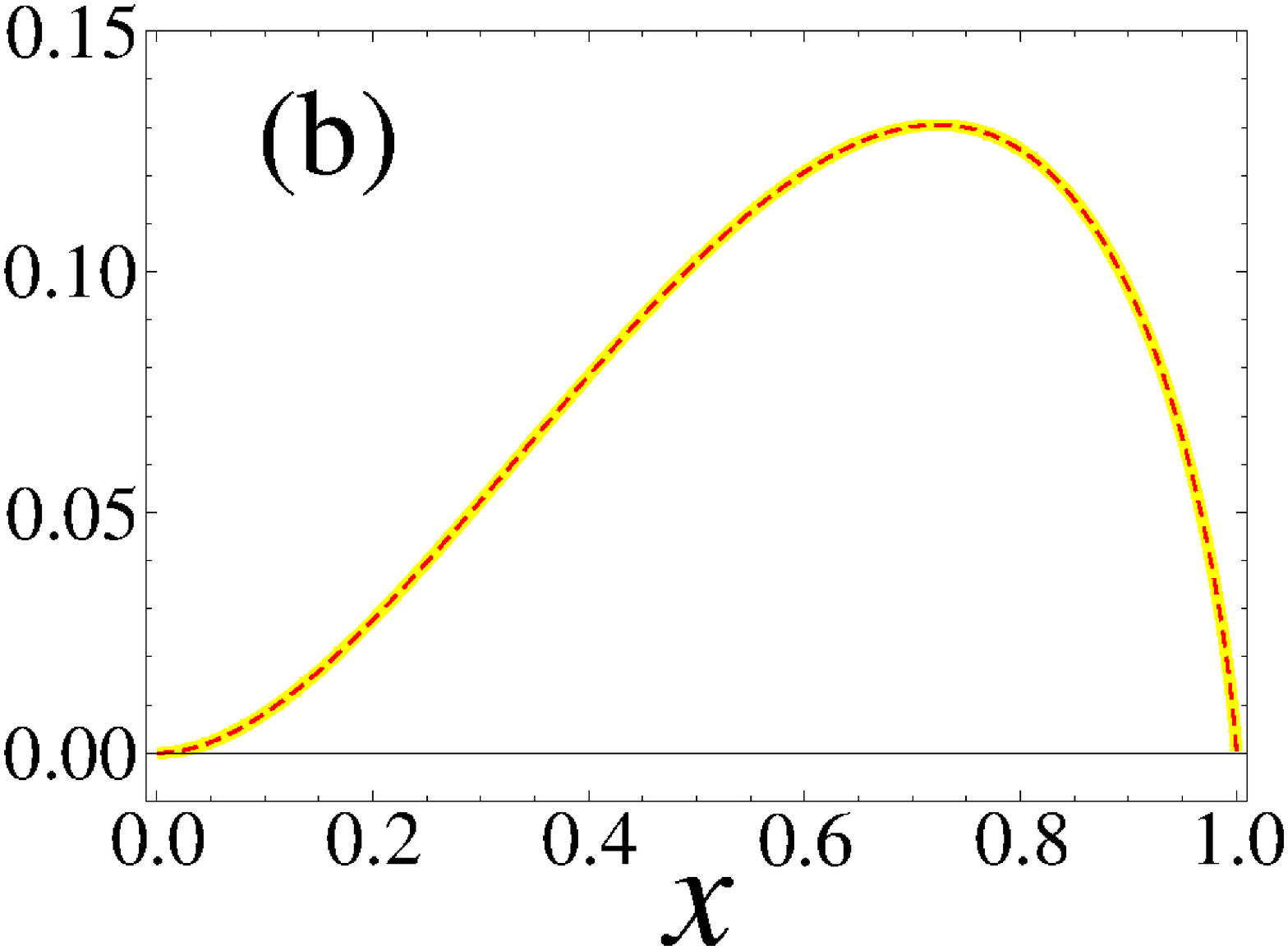}
\caption{(Color online) (a) $I_3/4$ (solid line) and $\mathfrak{I}$ (dashed line) as a function of $x$ for the state \eqref{rho}. Monogamy is violated by mixed states with $x\gtrsim 0.43596$. (b) $I_3-\mathfrak{I}$ (yellow thick line) and $\min_{(A,B,C)}(I_{A:C}+I_{B:C})$ (red dashed line) as a function of $x$ for the state \eqref{rho}. This simulation illustrates the validity of Eq.~\eqref{Ig+frakI}.}
\label{I3fig}
\end{figure}

Here we present some examples showing that monogamy is not always respected by mutual information. Consider the states~\cite{maloney13}
\be\begin{array}{l}
\rho_1=\frac{1}{4}\big(\ket{000}\bra{000}+\ket{011}\bra{011}+\ket{101}\bra{101}+\ket{110}\bra{110}\big), \\ \\
\rho_2=\rho_{AC}\otimes\rho_B, \\ \\
\rho_3=\frac{1}{2}\big(\ket{000}\bra{000}+\ket{111}\bra{111}\big).
\end{array}\nonumber \ee
By direct calculations one gets $\mathfrak{I}(\rho_1)=\ln 2$, $\mathfrak{I}(\rho_2)=0$, and $\mathfrak{I}(\rho_3)=-\ln 2$. For an illustration of the nontrivial behaviors of $I_3$ and $\mathfrak{I}$, we consider a three-qubit system in the state
\be\begin{array}{l}
\rho=\left(\tfrac{1-x}{8}\right)\mathbbm{1}_A\otimes\mathbbm{1}_B\otimes\mathbbm{1}_C+x\,\ket{\psi}\bra{\psi},\\ \\
\ket{\psi}=\alpha\,\ket{000}+\beta\,\ket{010}+\gamma\,\ket{101}+\delta\,\ket{111},
\end{array}\label{rho}\ee
where $x\in[0,1]$ and $|\alpha|^2+|\beta|^2+|\gamma|^2+|\delta|^2=1$.
Although an analytical formula for $\mathfrak{I}$ does exist for this state, it is not insightful and so it will be omitted. We then consider some particular cases. For $\alpha=\beta=\gamma=0$ we have that
\begin{subequations}
\be
I_3 &=& \tfrac{(1 + 7 x)}{8}\ln{(1 + 7 x)}-\tfrac{3(1 - x)}{8}\ln{(1 - x)}\nonumber \\ &-&\tfrac{4(1 + x)}{8} \ln{(1 + x)} -\tfrac{2(1 + 3 x)}{8} \ln{(1 + 3 x)}, \\ \nonumber \\
\mathfrak{I}&=&I_3+\tfrac{(1 - x)}{2} \ln{(1 - x)} +  \tfrac{4 (1 + x)}{2} \ln{(1 + x)} \nonumber \\ &-& \tfrac{(1 + 3 x)}{2} \ln{(1 + 3 x)}.
\ee
\end{subequations}
In Fig.~\ref{I3fig} the behavior of these quantities is shown as a function of $x$. For pure states $(x=1)$ one shows that $\mathfrak{I}=0$, as predicted by result 1 in Sec.~\ref{info}.

\end{document}